\documentclass[manuscript]{aastex}
\usepackage{txfonts}
\usepackage{graphicx}

\slugcomment{Submitted to ApJ}

\shorttitle{Extremely Energetic Outflow and Decelerated Expansion in W49N } \shortauthors{Liu et al.}

\begin{document}

\title{Extremely Energetic Outflow and Decelerated Expansion in W49N}
\author{Tie Liu\altaffilmark{1}, Kee-Tae Kim\altaffilmark{1}, Yuefang Wu\altaffilmark{2}, Di Li\altaffilmark{3}, Chang-Won Lee\altaffilmark{1}, Christopher G. De Pree\altaffilmark{4}, Sheng-Li Qin\altaffilmark{5}, Ke Wang\altaffilmark{6}, Ken'ichi Tatematsu\altaffilmark{7}, Qizhou Zhang\altaffilmark{8}, Diego Mardones\altaffilmark{9}, Sheng-Yuan Liu\altaffilmark{10}, Se-Hyung Cho\altaffilmark{1}   }
\altaffiltext{1}{Korea Astronomy and Space Science Institute 776, Daedeokdae-ro, Yuseong-gu, Daejeon, Republic of Korea 305-348; liutiepku@gmail.com}
\altaffiltext{2}{Department of Astronomy, Peking University, 100871, Beijing China}
\altaffiltext{3}{National Astronomical Observatories, Chinese Academy of Science, A20 Datun Road, Chaoyang District, Beijing 100012, China}
\altaffiltext{4}{Department of Physics and Astronomy, Agnes Scott College, Decatur, GA 30030, USA}
\altaffiltext{5}{Department of Astronomy, Yunnan University, and Key Laboratory of Astroparticle Physics of Yunnan Province, Kunming, 650091, China}
\altaffiltext{6}{European Southern Observatory, Karl-Schwarzschild-Str. 2, D-85748 Garching bei M\"{u}nchen, Germany}
\altaffiltext{7}{National Astronomical Observatory of Japan, 2-21-1 Osawa, Mitaka, Tokyo 181-8588, Japan}
\altaffiltext{8}{Harvard-Smithsonian Center for Astrophysics, 60 Garden Street, Cambridge, MA 02138, USA}
\altaffiltext{9}{Departamento de Astronom\'{\i}a, Universidad de Chile, Casilla 36-D, Santiago, Chile}
\altaffiltext{10}{Institute of Astronomy and Astrophysics, Academia Sinica, Taipei, Taiwan}

\begin{abstract}
W49N is a mini-starburst in the Milky Way and thus an ideal laboratory for high-mass star formation studies. Due to its large distance (11.1$_{-0.7}^{+0.9}$ kpc), the kinematics inside and between the dense molecular clumps in W49N are far from well understood. The SMA observations resolved the continuum emission into two clumps. The molecular line observation of SO$_{2}$ (28$_{4,24}$-28$_{3,25}$) suggests that the two clumps have a velocity difference of $\sim$7 km~s$^{-1}$. The eastern clump is very close to two radio sources ``G1" and ``G2", and the western clump coincides with a radio source ``B". The HCN (3-2) line reveals an extremely energetic outflow, which is among the most energetic molecular outflows in the Milky Way. This is the first report of high-velocity molecular outflow detection in W49N. The outflow jet might be in precession, which could account for the distribution, velocity and rotation of water maser spots. Three absorption systems are identified in HCO$^{+}$ (3-2) spectra. The absorption features are blueshifted with respect to the emission of SO$_{2}$ (28$_{4,24}$-28$_{3,25}$) lines, indicating that a cold layer is expanding in front of the warm gas. Further analysis indicates that the expansion is decelerated from the geometric expansion centers.
\end{abstract}


\section{Introduction}
High-mass stars play a major role in the evolution of galaxies. When compared with their low-mass counterparts, the outflow parameters (e.g. masses, energy and momentum) and mass infall rates in high-mass star formation regions are orders of magnitude larger \citep{wu04,beu02,zhang01,zhang05,qiu08,liu13b}, \textbf{indicating that high-mass star formation is a more energetic process.} Additionally, in contrast to isolated low-mass star formation, most of high-mass stars form in clusters. However, the formation of high-mass stars is still far from clearly understood due to difficulties in observations caused by their large distances, clustered environments, short evolutionary timescales and contamination from the feedback of protostars \citep{zin07}. To study physical and chemical environments of high-mass star forming regions, high spatial resolution interferometric observations are required.

Located at a distance of 11.1$_{-0.7}^{+0.9}$ kpc \citep{zhang13}, W49 is among the most luminous ($>10^{7}$ L$_{\sun}$) and massive ($\sim10^{6}$ M$_{\sun}$) star forming regions in the Milky Way \citep{si91,gal13}, making it an ideal laboratory for high-mass star formation studies. As shown in the left panel of Figure 1, W49 contains three massive star forming regions, W49 north (W49N), W49 south (W49S) and W49 southwest (W49SW), among which W49N is the most prominent one containing \textbf{40-50 of UC H{\sc ii} regions} which are arranged in a remarkable
2 pc diameter ring (``Welch ring") \citep{wel87,de97,smith09}. The ``Welch ring" is shown in the Upper-right panel of Figure 1.

Several scenarios have been proposed to explain the triggering mechanism of the mini-starburst in W49N, including global collapse \citep{wel87,will04,gal13}, cloud-cloud collision \citep{muf77,miya86,tar86,ser93,buck96,de97} and expanding shells \citep{peng10}. The global collapse scenario was proposed based on the detection of redshifted absorption features (``inverse P-Cygni profile") in HCO$^{+}$ (1-0) \citep{wel87} and CS (2-1) \citep{will04} lines. The Cloud-Cloud collision scenario is mainly based on the presence of overlapping clumps observed in molecular lines at different radial velocities \citep{ser93,buck96,de97}. \cite{peng10} identified two expanding shells in the mid-infrared images and confirmed by follow-up molecular line observations. They argued that the expanding shells in W49N are suggestive of triggered massive star formation in just $\sim$10$^{5}$ yrs. \textbf{Recently, \cite{ga13} argued that the W49N starburst most likely formed from global gravitational contraction with localized collapse in a ``hub–filament" geometry through large mosaic mapping observations with the Submillimeter Array (SMA) in combination with single-dish observations. Previous works have indicated that W49N contains substructures and forms a cluster of OB stars.} \textbf{However, the kinematics and interactions of different components in this region are still far from well understood. In this paper, for the first time, we report the discovery of an extremely energetic outflow and localized expansion in the central region of W49N from the Submillimeter Array (SMA) observations with an angular resolution of $\sim2\arcsec.4$.}

\section{Observations}

The SMA dataset is from the SMA\footnote{The Submillimeter Array is a joint project between the Smithsonian
Astrophysical Observatory and the Academia Sinica Institute of Astronomy and
Astrophysics, and is funded by the Smithsonian Institution and the Academia
Sinica.}  archive \citep{ho04}. The observations of W49N were carried out in
2005 September in its compact configuration. The phase reference center was R.A.(J2000)~=~19$^{\rm h}$10$^{\rm m}$13.41$^{\rm s}$ and
DEC.(J2000)~=~$09\arcdeg06\arcmin14.30\arcsec$. The 345 GHz receivers were tuned to 265 GHz for the lower sideband (LSB)
and to 275~GHz for the upper sideband (USB). Each sideband has 3072 channels and a total bandwidth of $\sim$2 GHz. The frequency spacing
across the spectral band is 0.812~MHz or $\sim$0.9 km s$^{-1}$. In the observations, QSOs 3C454.3, 1751+096, and planet Uranus were observed for bandpass correction, antenna-based gain correction and flux-density calibration, respectively.

MIRIAD was employed for calibration and imaging \citep{sau95}. The 1.1 mm continuum data were acquired by
averaging all the line-free channels over both the upper
and lower spectral bands. MIRIAD task "selfcal" was employed to
perform self-calibration on the continuum data.
The gain solutions from the self-calibration were applied to the
line data.

\textbf{The primary beam of the SMA at 270 GHz is $\sim$40$\arcsec$. The uv distances range from 10 to 62 $k\lambda$.} The synthesized beam size and 1 $ \sigma$ rms of the continuum emission
are $2.77\arcsec\times2.02\arcsec$ (PA=88.9$\arcdeg$) and $\sim$20 mJy~beam$^{-1}$, respectively. We smoothed the spectral lines to a spectral resolution of 1 km~s$^{-1}$ and the corresponding 1$ \sigma$ rms for lines is $\sim$0.1 Jy~beam$^{-1}$ per channel.

\section{Results}
\subsection{1.1 mm continuum emission}
The 1.1 mm continuum image is shown in red contours and color image in the \textbf{lower-right} panel of Figure 1. Two clumps are revealed in the 1.1 mm continuum emission. The 1.1 mm continuum emission can be well fitted with two 2-dimensional Gaussian sources as shown in blue contours. The peak positions of the eastern and western clumps are (-0$\arcsec$.35$\pm$0$\arcsec$.09, -1$\arcsec$.13$\pm$0$\arcsec$.03) and (-4$\arcsec$.06$\pm$0$\arcsec$.13, -1$\arcsec$.09$\pm$0$\arcsec$.06) offset from the phase center, respectively. The total integrated flux and deconvolved size of the eastern clump are 4.40 Jy and 2$\arcsec.04\times0\arcsec.71$ (P.A.=-74.7$\arcdeg$), respectively. Assuming a distance of 11.1 kpc, the effective radius of the eastern clump which is defined as $\sqrt{ab}$, where a and b are the full width at
half-maximum (FWHM) deconvolved sizes of major and minor axes, is $\sim$0.065 pc. The western clump is not resolved by the SMA and appears to be a point source with a total integrated flux of 2.02 Jy. In the upper panel of Figure 2, we present the 3.6 cm continuum emission \citep{de97} as a color image and also label the previously discovered bright radio continuum sources. We find the eastern continuum emission peak is very close to two radio sources ``G1" and ``G2". The western continuum emission peak coincides with a radio source ``B".

\subsection{Line emission}

Dozens of molecular lines were detected in both sidebands and were identified with the Splatalogue database for astronomical spectroscopy. Here we report the results from SO$_{2}$ (28$_{4,24}$-28$_{3,25}$), HCN (3-2), and HCO$^{+}$ (3-2) lines.

\subsubsection{SO$_{2}$ (28$_{4,24}$-28$_{3,25}$) emission}

The integrated intensity from -10 to 30 km~s$^{-1}$  of SO$_{2}$ (28$_{4,24}$-28$_{3,25}$) is shown in gray contours in the upper panel and black contours in the lower panel of Figure 2, respectively. The emission of SO$_{2}$ (28$_{4,24}$-28$_{3,25}$) coincides with the 1.1 mm continuum emission very well. From its intensity weighted velocity map (moment 1) in the lower panel of Figure 2, one can see that there is a large velocity gradient from east to west. This large velocity gradient is more clearly seen from the position-velocity diagram of SO$_{2}$ (28$_{4,24}$-28$_{3,25}$) (Figure 3), which is cut along the black dashed line in the lower panel of Figure 2. The peak velocities of SO$_{2}$ (28$_{4,24}$-28$_{3,25}$) are estimated from Gaussian fits at the individual positions and are marked with blue crosses in the lower panel of Figure 3. The peak velocities of SO$_{2}$ (28$_{4,24}$-28$_{3,25}$) at the continuum emission peaks of the eastern and western clumps are 5.8$\pm$0.1 and 12.7$\pm$0.1 km~s$^{-1}$, respectively. The velocity gradient between these two clumps is $\sim$34.7 km~s$^{-1}$~pc$^{-1}$.

\subsubsection{ Extremely high velocity emission in HCN (3-2) lines}

As shown in the position-velocity diagram (upper panel of Figure 3) and averaged spectrum (upper panel of Figure 4), the HCN (3-2) line emission shows extremely high velocity emission. The terminal velocities of the blueshifted and redshifted emission are as high as $\sim$-70 and $\sim$90 km~s$^{-1}$, respectively. The redshifted high-velocity emission is contaminated by HCN v2=1 (3-2) line, located at around 50 km~s$^{-1}$ as marked by the dashed dark brown line in the Position-Velocity diagram. There is also an emission peak in the averaged spectrum of HCN (3-2) (upper panel of Figure 4) at around 50 km~s$^{-1}$ due to the contamination of HCN v2=1 (3-2) line. \textbf{To confirm the contamination of HCN v2=1 (3-2), We integrated emission from 49 to 51 km~s$^{-1}$ and present the integrated intensity in Figure 5 as color image and white contours. The spatial distribution of HCN v2=1 (3-2) emission, even though being contaminated by HCN (3-2) high velocity emission, is roughly consistent with 1.1 mm continuum but significantly different from outflow emission as shown in Figure 4. The integrated intensity map in Figure 5 also rules out the possibility of contamination emission from foreground clouds, which might show a very extended structure if exists.}

\textbf{\cite{rob11} observed the J=3-2 and 4-3 transitions of HCN with the JCMT toward W49N. However, due to poor angular resolution and sensitivity, they did not detect such high velocity wings. Their spectral can be well fitted with two gaussian components, indicating the existence of substructures.} \textbf{The HCN outflow discovered in this work should be the first report of high velocity outflow detection in W49N.} We divide the outflow emission into ``Low" and ``High" components. The velocity intervals for each component are shown in the second column of Table 1. The emission between 40 and 60 km~s$^{-1}$ is not included in further analysis due to the contamination of HCN v2=1 (3-2) line. The integrated intensity maps of the blueshifted and redshifted ``High" outflow emission are shown as blue and red contours, respectively, in the lower panel of Figure 4. The blueshifted and redshifted outflow emission overlaps and only displaces by 0$\arcsec$.7 ($\Delta\theta$) in projection, indicating the outflow axis is almost along the line of sight. The projection \textbf{deconvolved} angular size of minor axes of outflow lobes is about 2$\arcsec$ ($\theta$). The upper limit of the outflow inclination angle \textbf{with respect to the line of sight} is $\varphi=\frac{\Delta\theta}{\theta}=\frac{0.7}{2}\times\frac{180^\circ}{\pi}\simeq20^\circ$. The characteristic outflow velocity (V$_{char}$) is defined as $V_{char}=V_{flow}-V_{sys}$, where V$_{sys}$ is the systemic velocity (5.8 km~s$^{-1}$) of the eastern continuum clump and V$_{flow}$ is the intensity weighted velocity of high velocity emission corrected with the projection effect by assuming an outflow inclination angle of 20$\arcdeg$. The characteristic outflow velocity of each outflow component is presented in the third column of Table 1. The V$_{char}$ of the high velocity redshifted outflow (``High Red") component is as high as 71.5 km~s$^{-1}$, indicating that the outflow is extremely energetic. The effective emission radius of each outflow component shown in the fourth column of Table 1 is derived from its deconvolved size and also corrected with the projection effect. The effective radii of the emission areas of the ``Low" and ``High" outflow components are 0.1 and 0.2 pc, respectively. The outflow properties will be discussed in detail in section 4.1.

\subsubsection{Absorption in HCO$^{+}$ (3-2) lines}

Figure 6 presents spectra of HCO$^{+}$ (3-2) and SO$_{2}$ (28$_{4,24}$-28$_{3,25}$) at six positions. Three absorption systems in HCO$^{+}$ (3-2) lines are identified. The mean velocities of these three absorption systems are ``-7.5 km~s$^{-1}$", ``2.0 km~s$^{-1}$", and ``12.5 km~s$^{-1}$", respectively. These three absorption systems can also be clearly seen from the Position-Velocity diagrams in Figure 3 as in dashed contours. We made the integrated intensity maps in the velocity intervals of [-10,-4], [-2,6] and [10,15] km~s$^{-1}$ for the ``-7.5 km~s$^{-1}$", ``2.0 km~s$^{-1}$", and ``12.5 km~s$^{-1}$" absorption systems, respectively, and display them as dashed contours in the upper panel of Figure 2. The ``-7.5 km~s$^{-1}$" absorption is associated with radio continuum sources ``H" and ``G4". The ``-2.0 km~s$^{-1}$" absorption perfectly corresponds to the 3.6 cm continuum emission of ``G", which includes sources ``G1" to ``G5". The ``12.5 km~s$^{-1}$" absorption is associated with radio source ``B". Additionally, from the spectra in Figure 6 and Position-Velocity diagram in Figure 3, \textbf{one can see that HCO$^{+}$ (3-2) absorption dips are always at the blueshifted side of the SO$_{2}$ (28$_{4,24}$-28$_{3,25}$) emission except in the overlapped regions (e.g. (-1$\arcsec$.5,-1$\arcsec$.5) and (-2$\arcsec$.5,-1$\arcsec$.5) off positions) of the two clouds.} \textbf{The HCO$^{+}$ (3-2) spectra in the right column of Figure 5 have a typical ``P-Cygni" profile with emission peaks redshifted with respect to absorption dips.} The peak velocities of HCO$^{+}$ (3-2) absorption dips estimated from gaussian fit at each position are also marked with red crosses in the lower panel of Figure 3. One can see that the absorption dips of HCO$^{+}$ (3-2) are \textbf{mostly} blueshifted to the emission peaks of SO$_{2}$ (28$_{4,24}$-28$_{3,25}$). \textbf{Since SO$_{2}$ (28$_{4,24}$-28$_{3,25}$) (E$_{u}\sim416$K) has much higher upper energy level than that of HCO$^{+}$ (3-2) (E$_{u}\sim26$K), this velocity placement indicates that HCO$^{+}$ (3-2) absorption traces an \textbf{localized} expanding cold layer in front of the warm gas traced by SO$_{2}$ (28$_{4,24}$-28$_{3,25}$) emission.} More discussion about the absorption in HCO$^{+}$ (3-2) lines will be presented in section 4.2.

\section{Discussion}

\subsection{Extremely energetic molecular outflow}

\subsubsection{The outflow properties}

As presented in section 3.2.2, the HCN (3-2) lines show extremely high velocity emission, indicating energetic outflows. Following \cite{Liu11b}, the total mass of each outflow lobe is given by:
\begin{equation}
M = 1.04~\times~10^{-4}D^{2}\frac{Q_{rot}e^{E_{u}/T_{rot}}}{\chi\nu^{3}S\mu^{2}}\int\frac{\tau}{1-e^{-\tau}}S_{\nu}dv
\end{equation}
where M, D, S$_{\nu}$, $\chi$, and $\tau$ are the outflow
gas mass in M$_{\sun}$, source distance in kpc, line flux density in
Jy, fractional abundance \textbf{with respect to H$_{2}$}, and optical depth, respectively. Q$_{rot}$, E$_{u}$, T$_{rot}$, $\nu$, S, $\mu$  are the
rotational partition function, the upper level energy in
K, the rotation temperature, the rest frequency, line strength and the permanent
dipole moment, respectively. Owing to the lack of a direct estimation of the fractional abundance of HCN in
the outflow region of W49N, we adopt the same fractional abundance of $5.2\times10^{-9}$ as that in G9.62+0.19 F \citep{Liu11b} in calculating
the outflow parameters. Assuming that the high-velocity emission is optically thin and T$_{rot}$=30 K \citep{Liu11b}, the outflow
masses were calculated and presented in the sixth column of Table 1. The total outflow mass is \textbf{$\sim$40 M$_{\sun}$}. The momentum can be calculated by
$P=M\times V_{char}$, and the energy by $E=\frac{1}{2}M\times V_{char}^{2}$. The momentum and energy of the outflows are shown in the seventh and eighth columns, respectively. The total momentum and energy are
\textbf{1700 M$_{\sun}$~km~s$^{-1}$} and $8.7\times10^{47}$ erg, respectively. The dynamical timescale t is estimated as 2R$_{eff}$/V$_{char}$ and presented in the fifth column of Table 1. The dynamic
timescale is about $0.5-1\times10^{4}$ year. In comparison with surveys of molecular outflows in massive star forming regions \citep{zhang01,zhang05}, the outflow mass, momentum and energy in W49N are in the upper bound of the outflow parameters of massive star formation.

The mechanical luminosity L$_{m}$, and the mass-loss rate $\dot{M}$ are
calculated as L$_{m}$=E/t and $\dot{M} = P/(tV_{w})$. \textbf{Here we assume that the outflow is powered by winds driven by accretion disks and the wind velocity $V_{w}$ is 500 km~$^{-1}$ \citep{zhang05}.} The mechanical luminosity and the mass-loss rates are listed in the last two columns of Table 1. The total mechanical luminosity and the mass-loss rate of the outflows are estimated
to be 1.3$\times10^{3}$ L$_{\sun}$ and $5.6\times10^{-4}$~M$_{\sun}$$\cdot$yr$^{-1}$, respectively. Assuming that only one third of accreted mass is ejected in winds \citep{shu87}, we find a
lower limit to the accretion rate to be $1.7\times10^{-3}$~M$_{\sun}$$\cdot$yr$^{-1}$.

\cite{smith09} identified a mid-IR source (G:IRS1) and assigned it to the driving source of the outflow. From SED modeling, they inferred that the embedded protostar has a stellar mass of $\sim$45 M$_{\sun}$, luminosity $\sim3\times10^{5}$L$_{\sun}$ and an equivalent envelope accretion rate $\sim10^{-3}$M$_{\sun}$~yr$^{-1}$. The mechanical luminosity of O-type stars with a mass loss rate of $\dot{M}$ and stellar wind velocity $v_{w}$ can be estimated from \citep{Nagy12}:
\begin{equation}
L_{w}=1.3\times10^{36}(\frac{\dot{M}}{10^{-6}M_{\sun}~yr^{-1}})(\frac{v_{w}}{2\times10^{3}km~s^{-1}})^{2} erg~s^{-1}
\end{equation}
The mechanical luminosity of an O type star with a mass loss rate of $10^{-6}$M$_{\sun}$~yr$^{-1}$ and stellar wind velocity of 2000 km~s$^{-1}$ \citep{smith06} is $\sim$340 L$_{\sun}$. Assuming an efficiency of 10\%, its mechanical heating rate (34 L$_{\sun}$) is much smaller than the mechanical luminosity (1.3$\times10^{3}$ L$_{\sun}$) of outflows, indicating that the outflow could not be driven by the stellar winds of a single O type star (e.g. G:IRS1). The outflow might be driven by accretion as suggested by \cite{smith09}. \textbf{The envelope accretion rate is in their model fits to the IR SED is $1\times10^{-3}$~M$_{\sun}$$\cdot$yr$^{-1}$ \citep{smith09}, which is consistent with the accretion rate ($1.7\times10^{-3}$~M$_{\sun}$$\cdot$yr$^{-1}$) inferred from the HCN outflow.}

There are large uncertainties in the estimation of the outflow parameters. The optically thin assumption may not be true for low-velocity outflows. Assuming an average optical depth of 1.5, the corresponding outflow parameters (M, P, E, $\dot{M}$ and L$_{m}$) will become two times larger. As mentioned before, we ignored the emission at velocities between 40 to 60 km~s$^{-1}$ due to the contamination of HCN v2=1 (3-2) line, which definitely leads to underestimation of the outflow parameters for redshifted high-velocity emission. The outflow parameters are less sensitive to the temperature. If we take a rotation temperature of 60 K (two times larger than the value we used), the outflow parameters only become 30\% smaller. The largest uncertainties come from the fractional abundance of HCN and inclination angle.\textbf{ \cite{rob11} derived an H$_{2}$ volume density n$_{H_{2}}$ of 2$\times10^{6}$ cm$^{-3}$ and an HCN column density of (2-7)$\times10^{15}$ cm$^{-2}$ in the ``Core" region of W49N. The angular size of HC$^{15}$N (4-3) emission is $L\sim23\arcsec$. We can estimate an average H$_{2}$ column density: $N_{H_{2}}=L\times N_{H_{2}}=7.6\times10^{24}$ cm$^{-2}$. Therefore, the fraction abundance of HCN at ``Core" region is $(2.6-9.2)\times10^{-10}$, which is about one order of magnitude smaller than the value ($5.2\times10^{-9}$) we used, indicating that the outflow parameters could be significantly underestimated.} However, HCN appears greatly enhanced in some outflow regions. In NGC 1333 IRAS 2A outflow region, the fractional abundance of HCN is about twice larger than that in the core region \citep{jor04}. While in L1157 outflow region, the HCN abundance is even enhanced by a factor of 100. Thus, more accurate determination of the HCN abundance in outflow regions is critically important in estimation of outflow parameters. \cite{su07} estimated an HCN abundance of $1-2\times10^{-8}$ in the massive outflow lobes of IRAS 20126+4104. If we assume that HCN fractional abundance in W49N is 1$\times10^{-8}$ as in IRAS 20126+4104 outflows, the outflow parameters will become about two times smaller. The inclination angle directly affects the estimation of R$_{eff}$ and V$_{char}$. The outflow parameters of M, P, and E do not change too much ($<10\%$) when the inclination angle changes from 10$\arcdeg$ to 30$\arcdeg$. While the corresponding $\dot{M}$ and L$_{m}$ can change by a factor of 2. In general, the respective uncertainty of outflow parameters due to the effects discussed above \textbf{could be at least one order of magnitude}. \textbf{The outflow in W49N was not detected in CO (2-1) line emission with an angular resolution of 3$\arcsec$ \citep{ga13}. The reason may be that the outflow is very compact and warm, which cannot be revealed with low excitation lines. Further higher angular resolution and higher sensitivity observations of multiple higher excitation transitions of HCN and other molecular lines (e.g. CO, SiO) may help to better constrain the outflow properties.}

HCN outflows have been detected by interferometers toward several other high-mass star forming regions \citep{su07,liu10,Liu11a,Liu11b,zhu11,liu13a}. Comparing with those outflows, W49N has much larger outflow velocity. Among the previous detected HCN outflows, the one in G9.62+0.19 F might be the most energetic one \citep{Liu11b}. The outflow in W49N has similar mass and \textbf{momentum }to those of G9.62+0.19 F, but the energy and mechanical luminosity are about 10 times larger. \cite{wu04} cataloged 391 outflows, most of which were detected by single dishes. When compared with those outflows, the one in W49N is also among the most luminous and energetic outflows. The maximum energy and mechanical luminosity of the outflows in \cite{wu04} are around 10$^{48}$ erg and 10$^{3}$ L$_{\sun}$, respectively, which are comparable to that in W49N. Thus, we argue that the outflow in W49N is among the most energetic outflows in the Milky Way. \textbf{However, we should caution that the HCN outflow is barely resolved by the SMA and may contain multiple outflows as seen in other high-mass cluster forming regions (e.g. NGC7538S; \cite{nar12}). Therefore, the outflow parameters estimated here may be the sum of multiple outflows. }

\subsubsection{Does the outflow excite the water masers?}

W49N harbors the most luminous water maser activities in the Milky Way \citep{walk82}. The water masers in W49N are mainly located near the H{\sc ii} regions ``G1" and ``G2" \citep{mor73,walk82,gw92,de00,mcg04}. \cite{mcg04} identified 316 water maser components with a velocity spanning from -352.1 to 375.5 km~s$^{-1}$. In the lower panel of Figure 4, we overlay the water maser spots on the integrated intensity contours of HCN (3-2) high-velocity outflow emission. The water maser spots near ``G1" and ``G2" are distributed in three distinct clusters. Those maser spots with blueshifted and redshifted high-velocity features are located to the west and to the east of the mid-IR source ``G:IRS1", respectively. The tightest cluster of maser spots is close to ``G:IRS1" and has a linear arrangement elongated along a north-south axis. \cite{hon04} detected a strong maser outburst on an arc-like structure in the central cluster, which may be due to a shock phenomenon powered by a forming star or a star cluster. Recently, from proper motion measurements, \cite{zhang13} found that those water maser spots appear to be in expansion and rotation with expanding and rotating velocities of 10.99$\pm$1.33 and 5.28$\pm$0.80 km~s$^{-1}$.

The water maser emission in W49N has been a puzzle to astronomers for decades. \cite{tar86} suggested a cloud-cloud collision model for water maser excitation in W49N. An alternative promising explanation is that water masers form
at the edge of the outflow cavity walls (or jet-driven cocoons) \citep{mac94,smith09}. Such model for water maser excitation requires a shock velocity on the cavity walls of $\sim$20-200 km~s$^{-1}$. Since the outflow velocity of HCN (3-2) is up to $\sim$70 km~s$^{-1}$, water masers should be easily excited at the edge of the outflow cavity walls. Additionally, the low inclination angle of the outflow might account for the observed high velocity features of water masers. As mentioned before, the blueshifted and redshifted high-velocity emission of HCN (3-2) overlaps and only displaces by $\sim0\arcsec.7$ in projection, which is not resolved by the SMA. However, the orientation of the HCN outflow is roughly consistent with the water maser outflow. The western cluster of water masers is located at the peak of the blueshifted high-velocity HCN emission. While the eastern cluster of water maser emission is mainly at the red lobe of HCN outflow. A $\sim$25 km~s$^{-1}$ CO outflow was observed by \cite{sco86}, where the blueshifted CO emission
is towards the east and the redshifted CO emission is to the west. The orientation for the low-velocity CO outflow is opposite to the high-velocity HCN outflow, indicating that the outflow jet \textbf{might be} in precession. The jet precession naturally explains why the high- and low velocity features coexist in the centeral cluster of water masers, while at the western and eastern clusters of waters masers, only high-velocity features are observed \citep{mcg04}. \textbf{The expansion of the water maser spots might be caused by outflows. The jet precession can naturally explain why the maser spots are also in rotation in spite of expansion.} \textbf{However, we should point out that the jet precession scenario needs to be tested by higher angular resolution observations because: 1. The outflow in W49N is not obvious in new CO (2-1) line observations with the SMA \cite{ga13}; 2. The HCN outflow is barely resolved by the SMA and may contain multiple outflows. }

\subsection{Decelerated expansion?}

As discussed in section 3.2.3, the absorption dips of  HCO$^{+}$ (3-2) are all blueshifted with respect to the emission peaks of SO$_{2}$ (28$_{4,24}$-28$_{3,25}$), indicating that the cold layers may be in expansion. Since there is no SO$_{2}$ (28$_{4,24}$-28$_{3,25}$) emission corresponding to the ``-7.5 km~s$^{-1}$" absorption system, here we only focus on the ``2.0 km~s$^{-1}$" and ``12.5 km~s$^{-1}$" absorption systems. The systemic velocities of the warm layers can be well determined from SO$_{2}$ (28$_{4,24}$-28$_{3,25}$) emission lines. We define the line-of-sight (los) expansion velocity V$_{exp}^{los}$ as V$_{exp}^{los}$=V$_{emi}$-V$_{abs}$, where V$_{emi}$ is the velocity of emission peak of SO$_{2}$ (28$_{4,24}$-28$_{3,25}$) and V$_{abs}$ is the velocity of absorption dip of HCO$^{+}$ (3-2). \textbf{V$_{emi}$ and V$_{abs}$ were derived from gaussian fits and marked with blue and red crosses in the Position-Velocity diagram in the lower panel of Figure 3, respectively.} We derive V$_{exp}^{los}$ at different positions along the black dashed line in the lower panel of Figure 2 and plot V$_{exp}^{los}$ versus the RA offsets in the upper panel of Figure 7. \textbf{We ignored the positions where the ``2.0 km~s$^{-1}$" and ``12.5 km~s$^{-1}$" absorption systems overlaps such as off positions (-1$\arcsec$.5,-1$\arcsec$.5) and (-2$\arcsec$.5,-1$\arcsec$.5).}

\textbf{There are two peaks in the ``V$_{exp}^{los}$ vs. RA" plot in the upper panel of Figure 7:} the eastern one around 2$\arcsec$.5 and the western one around -6$\arcsec$. This ``V$_{exp}^{los}$ vs. RA" curve can not be simply fitted with a spherically expanding model with a constant expansion velocity, in which the Position-Velocity curve would be elliptical like. Instead, we propose a toy model, which is shown in the lower panel of Figure 7, to explain the absorption of HCO$^{+}$ (3-2). Relative to the warm gas layer which is distributed in the central plane, the two parallel elliptical cold layers are moving outward. The positions of the cold layer can be described with two parameters: height from the central plane (``h") and distance (``r") to the geometric expansion center ``O" whose RA offset is ``X0". The maximum of ``h" and ``r" are ``b" and ``a" respectively. Assuming this simple geometric model, we find the ``V$_{exp}^{los}$ vs. RA" curve can be well fitted with V$_{exp}\propto$~r$^{-1}$, indicating that the expansion decreases with the distance to the geometric expansion center. \textbf{V$_{exp}$ is the real expanding velocity: V$_{exp}=\frac{V_{exp}^{los}}{\cos(\theta)}$, where $\theta$ is the angle of velocity vector with respect to the line of sight.} The maximum  expansion velocity V$_{max}$ corresponds to the minimum distance ``b". \textbf{ The best fits with smallest $\chi^{2}$} are shown in red and blue solid lines in the upper panel of Figure 7. The parameters [b, a, X0, V$_{max}$] of the best fits to the western and eastern expansion systems are [1.18$\arcsec$, 2.84$\arcsec$, -6.24$\arcsec$, 3.37 km~s$^{-1}$] and [1.43$\arcsec$, 4.57$\arcsec$, 2.52$\arcsec$, 5.27 km~s$^{-1}$], respectively.

The dynamical timescale can be estimated as t$_{exp}=\frac{b}{V_{max}}$. The dynamical timescale for the western and eastern expansion systems are 1.8$\times10^{4}$ and 1.4$\times10^{4}$ yr, respectively. The actual masses (M) of the expanding cloud can be roughly estimated from their virial masses M$_{vir}$ \citep{kau13}:
\begin{equation}
\frac{M}{10^{3}M_{\sun}}=M_{vir}/\alpha_{cr}=1.2(\frac{\sigma_{v}}{km~s^{-1}})^{2}(\frac{R_{eff}}{pc})/\alpha_{cr}
\end{equation}
where $\alpha_{cr}$ is the so-called critical virial parameter and is found to be 2$\pm$1 for a wide range of cloud shapes and density gradients. \textbf{Here we take $\alpha_{cr}$ of 2 \citep{kau13}.} The radius R$_{eff}$ here is taken as a effective radius of $\sqrt{ab}$. The R$_{eff}$ of the western and eastern expanding clouds are $\sim$0.10 and $\sim$0.14 pc,respectively. Velocity dispersion $\sigma_{v}$ is measured from the linewidth of the absorption lines of HCO$^{+}$ (3-2). The $\sigma_{v}$ is found to be 1.5 and 1.8 km~s$^{-1}$ for the western and eastern expanding clouds, respectively. Thus, the masses of the western and eastern expanding clouds are $\sim$270 and $\sim$130 M$_{\sun}$, respectively. For ``subcritical" clouds, which are unbound and expanding, the virial parameter should be larger than the critical virial parameter. Thus, the actual mass estimated from equation (3) should be taken as an upper limit. We define an effective expanding velocity as V$_{eff}$=$\frac{b}{R_{eff}}\times V_{max}$. V$_{eff}$ of the western and eastern expanding clouds are $\sim$2.2 and $\sim$2.9 km$^{-1}$, respectively. Then we can estimate the momentum and kinetic energy of the expanding clouds as
P=M$\times$V$_{eff}$ and E$_{kin}=\frac{1}{2}$M$\times$V$_{eff}^{2}$, respectively. The momentum of the western and eastern expanding clouds are $\sim$290 and 790 M$_{\sun}$~km~s$^{-1}$, respectively. The kinetic energy of the western and eastern expanding clouds are $\sim6.3\times10^{45}$ and $\sim2.3\times10^{46}$ erg, respectively. The corresponding mechanical luminosity L$_{m}$ is calculated as L$_{m}$=E$_{kin}$/t$_{exp}$. The L$_{m}$ of the western and eastern expanding clouds are 2.9 and 13.7 L$_{\sun}$, respectively.

\cite{smith09} suggests that the cavity encircled by radio source ``G1" to ``G5" is caused by the outflow. Assuming an efficiency of 1-10\%, the mechanical heating rate of the outflow is about 13 to 130 L$_{\sun}$, which is comparable or much larger than the mechanical luminosity of the eastern expanding cloud, indicating that the outflow has sufficient energy to blow the cavity. The extent and timescale of the outflow are also comparable to those of the eastern expanding cloud which is associated with the cavity. All indicate that the cavity is formed due to the outflow. \textbf{However, this scenario is doubtful because that the outflow appears to have a very small inclination angle and the outflow lobes are very compact (0.1-0.2 pc in radius). The red outflow lobe is about 2.7$\arcsec$ ($\sim$0.15 pc) away from the expanding center of the cavity. This displacement is comparable to the radii of outflow lobes and could be resolved by present SMA beam size ($\sim$2.4$\arcsec$).} On the other hand, we also notice that the stellar wind of an O-type star also has sufficient energy to create such cavity.  The mechanical luminosity of an O type star with a mass loss rate of $10^{-6}$M$_{\sun}$~yr$^{-1}$ and stellar wind velocity of 2000 km~s$^{-1}$ is $\sim$340 L$_{\sun}$. Assuming an efficiency of 10\%, its mechanical heating rate is larger than the mechanical luminosity (13.7 L$_{\sun}$) of the eastern expanding cloud. Therefore, we suggest that the expansion of the cavity is more likely due to the mechanical heating produced by the stellar winds rather than outflow. The western expanding cloud is associated with a radio source ``B", which contains one O6.5 and one O5.5 ZAMS type stars \citep{de00}. The HCO$^{+}$ lines show typical ``P-Cygni" profile toward radio source ``B", which indicates the expansion of an H{\sc ii} region.

\textbf{Two expanding shells with an average radius of $\sim$3.3 pc have been identified the mid-infrared images and the $^{13}$CO J=2-1 and C$^{18}$O J=2-1 data from the IRAM 30-m observations \citep{peng10}. The average mass of the expanding shells is $2\times10^{4}$ M$_{\sun}$ with a kinetic energy of $\sim10^{49}$ erg, which are 2 or 3 orders of magnitude larger than the expanding systems reported in this paper. The expansion traced by HCO$^{+}$ (3-2) line is more localized when compared with those infrared expanding shells. The connection between HCO$^{+}$ expansion systems and infrared expanding shells are not clear. }

\section{Summary}
We have studied W49N with the SMA data in the 1.1 mm continuum and molecular line emission. The main results of this study are as follows:

1. Two clumps are revealed in the 1.1 mm continuum emission. The eastern one is very close to radio source ``G1" and ``G2", and the western clump coincides with the radio source ``B".

2. HCN (3-2) lines show extremely high-velocity emission, indicating energetic outflow motions. \textbf{This is the first report of high-velocity outflow detection in W49N.} The total outflow mass is 40 M$_{\sun}$. The total momentum and energy are
1700 M$_{\sun}$~km~s$^{-1}$ and $8.7\times10^{47}$ erg, respectively. The dynamic
timescale of the outflow is about $0.5-1\times10^{4}$ year.  The total mechanical luminosity and the mass-loss rate are estimated
to be 1.3$\times10^{3}$ L$_{\sun}$ and \textbf{$5.6\times10^{-4}$~M$_{\sun}$$\cdot$yr$^{-1}$, respectively. The corresponding accretion rate is at leat $1.7\times10^{-3}$~M$_{\sun}$$\cdot$yr$^{-1}$.} The orientation for the high-velocity HCN outflow is opposite to the low-velocity CO outflow, indicating that the outflow jet \textbf{might be} in precession. The jet precession can naturally explain the distribution, velocity and rotation of water maser spots.

3. Three absorption systems are identified in HCO$^{+}$ (3-2) spectra, which are at ``-7.5 km~s$^{-1}$", ``2.0 km~s$^{-1}$", and ``12.5 km~s$^{-1}$", respectively. The ``-7.5 km~s$^{-1}$" absorption is associated with radio continuum sources ``H" and ``G4". The ``2.0 km~s$^{-1}$" absorption perfectly corresponds to the 3.6 cm continuum emission of ``G", which includes sources ``G1" to ``G5". The ``12.5 km~s$^{-1}$" absorption is associated with radio source ``B". The absorption of HCO$^{+}$ (3-2) is blueshifted with respect to the emission of SO$_{2}$ (28$_{4,24}$-28$_{3,25}$) lines, indicating a cold layer is expanding in front of the warm gas. Analysis toward the ``2.0 km~s$^{-1}$", and ``12.5 km~s$^{-1}$" absorption system indicates that the expansion velocity linearly decreases with the distance to the geometric expansion center. The expansion traced by HCO$^{+}$ (3-2) absorption features might be caused by mechanical heating of stellar winds from associated O type stars.

\section*{Acknowledgment}
\begin{acknowledgements}
We are grateful to the \textbf{SMA archive staff}. Tie Liu is supported by KASI fellowship. S.-L. Qin is partly supported by NSFC under grant Nos. 11373026, 11433004, by Top Talents Program of Yunnan Province. Ke Wang acknowledges support from the ESO fellowship. Y. Wu is partly supported by the China Ministry of Science and
Technology under State Key Development Program for Basic Research
(No.2012CB821800), the grants of NSFC No.11373009 and No.11433008.

\end{acknowledgements}

\begin{deluxetable}{ccrrrrrrrrrrrrrcrl}
\tabletypesize{\scriptsize} \tablecolumns{15} \tablewidth{0pc}
\tablecaption{Outflow parameters derived from HCN (3-2) lines} \tablehead{
\colhead{Component} & \colhead{Velocity interval} & \colhead{V$_{char}$\tablenotemark{a} } & \colhead{R$_{eff}$\tablenotemark{a}} &
\colhead{t} & \colhead{M} &
\colhead{P} & \colhead{E}  & \colhead{$\dot{M}$} & \colhead{L$_{m}$} \\
\colhead{} & \colhead{(km~s$^{-1}$)} &
\colhead{(km~s$^{-1}$)} &
\colhead{(pc)} &
\colhead{($10^{3}$yr)} &
\colhead{(M$_{\sun}$)} &
\colhead{(M$_{\sun}\cdot$km~s$^{-1}$)} & \colhead{($10^{47}$erg)} & \colhead{(10$^{-4}$M$_{\sun}$~yr$^{-1}$)} & \colhead{(10$^{2}$L$_{\sun}$)}  } \startdata
Low Blue  &[-30,-10]  &    26.2 &     0.1  &     10.9 & 12     &   300   &  0.8    & 0.6 &  0.6  \\
Low Red   &[30,40]    &    29.5 &     0.1  &     7.8 & 12     &   350   &  1.0    & 0.9 &  1.1      \\
High Blue &[-80,-30]  &    55.7 &     0.2  &     6.2 &  6     &   340   &  1.9    & 1.1 &  2.6  \\
High Red  &[60,120]   &    71.5 &     0.2  &     4.8 &  10     &   710   &  5.0    & 3.0 &  8.7  \\
\enddata
\tablenotetext{a}{corrected with an inclination angle of 20$\arcdeg$ \textbf{with respect to the line of sight}. \textbf{R$_{eff}$ is the full width at half-maximum (FWHM) deconvolved
size.} }
\end{deluxetable}

\begin{figure}
\centering
\includegraphics[angle=-90,scale=0.6]{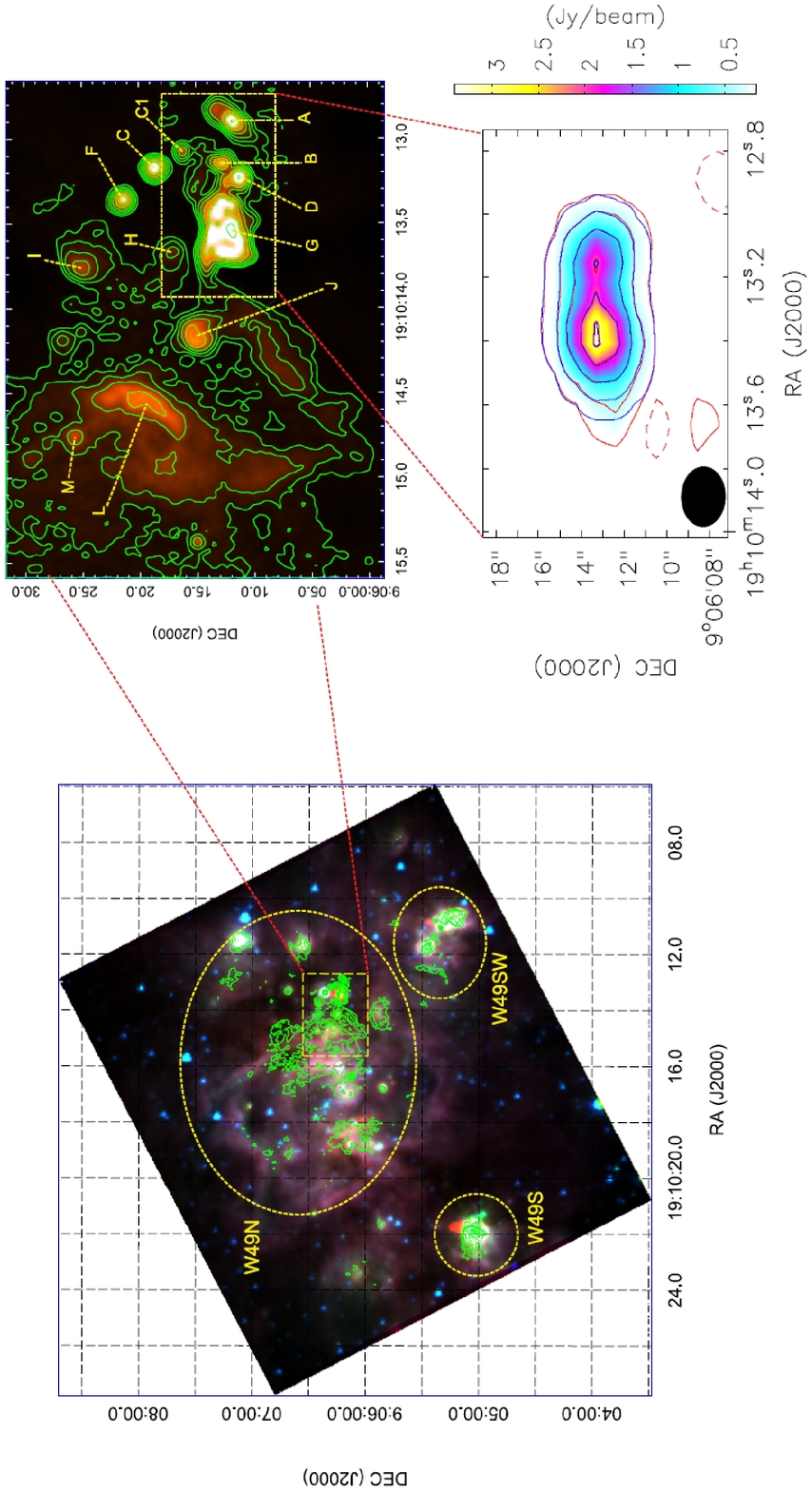}
\caption{Left: The 3.6 cm continuum emission \citep{de97} is shown in contours overlayed on the 3 color composite image from Spitzer/IRAC (3.6 $\micron$ in blue, 4.5 $\micron$ in green and 8.0 $\micron$ in red). The contours start at the 3$\sigma$ level (1.0 mJy~beam$^{-1}$). Subsequent positive contours are at 2, 4, 8, 16, 32 times the 3$\sigma$ level. Upper-right: The ring of UCH{\sc ii} regions (``Welch ring"). The color image and contours represent the 3.6 cm continuum emission. The names of strongest UCH{\sc ii} regions are labeled. Lower-right: The 1.1 mm continuum emission observed by the SMA is shown in color image and red contours. The blue contours represent the 2-dimensional gaussian fits. The contours are (5,20,50,100,150)$\times$0.02 Jy~beam$^{-1}$ (1 $\sigma$). The SMA beam is shown in filled black ellipse. }
\end{figure}

\begin{figure}
\centering
\includegraphics[angle=0,scale=0.4]{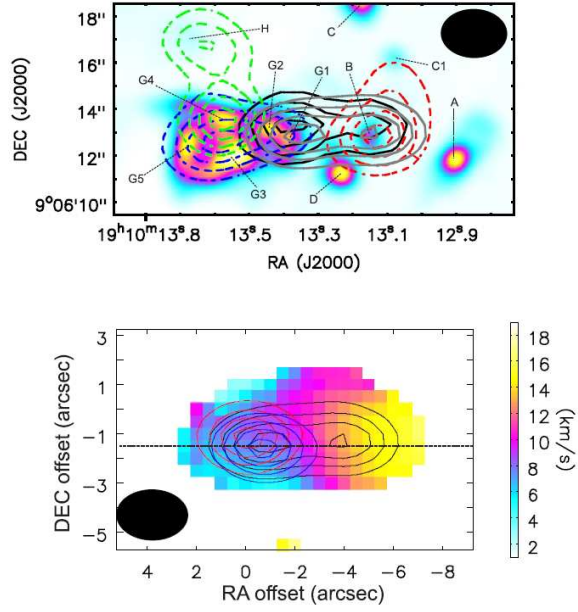}
\caption{Upper: The 3.6 cm continuum emission \citep{de97} is shown in color image. The names of bright cm continuum sources are labeled from ``A" to ``H". The integrated intensity maps of ``-7.5 km~s$^{-1}$", ``2.0 km~s$^{-1}$", and ``12.5 km~s$^{-1}$" absorption of HCO$^{+}$ (3-2) are shown in green, blue and red dashed contours, respectively. The 1.1 mm continuum emission and integrated intensity map of SO$_{2}$ (28$_{4,24}$-28$_{3,25}$) are shown in solid black and gray contours, respectively. All the contours are started from 30\% in step of 20\% of the peak values. The peak values for the green, blue, red, gray and black contours are -12.8, -19.8, -8.7, 106.2 Jy~beam$^{-1}$~km~s$^{-1}$ and 3.2 Jy~beam$^{-1}$, respectively. Lower: The intensity weighted velocity (Moment 1) map of SO$_{2}$ (28$_{4,24}$-28$_{3,25}$) is shown in color image. The integrated intensity map of SO$_{2}$ (28$_{4,24}$-28$_{3,25}$) is shown in black contours. The contours are [-3, 3, 5, 10, 20, 40, 80, 160, 240, 320]$\times$0.3 Jy~beam~km~s$^{-1}$ \textbf{(3 $\sigma$)}. The 1.1 mm continuum emission is shown in blue contours. The SMA beam is shown in filled black ellipse. }
\end{figure}

\begin{figure}
\centering
\includegraphics[angle=0,width=100mm]{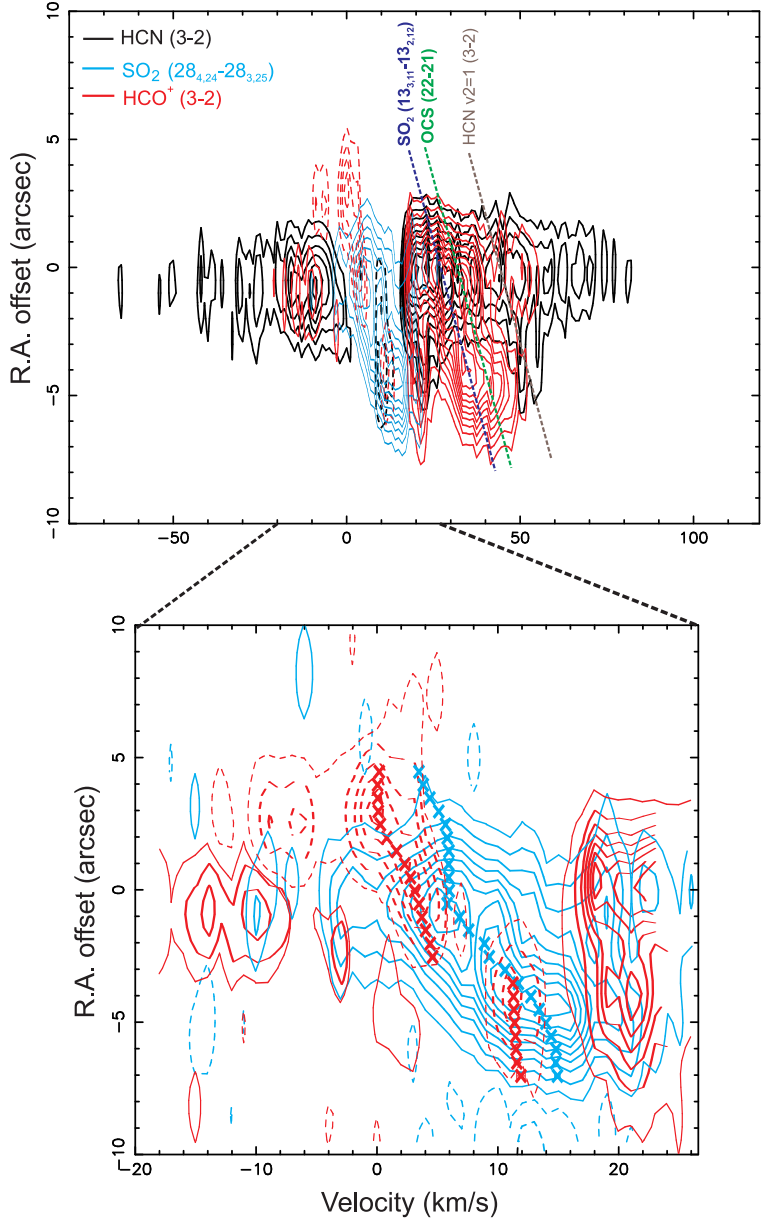}
\caption{Position-Velocity diagrams of HCN (4-3), SO$_{2}$ (28$_{4,24}$-28$_{3,25}$) and HCO$^{+}$ (3-2) are shown in black, blue and red contours, respectively. The Position-Velocity diagrams are made by cutting along the black dashed line in Figure 2. The R.A. offsets are relative to 19$^{\rm h}$10$^{\rm m}$13.41$^{\rm s}$. All the contours in the upper panel are started from 1 Jy~beam$^{-1}$ (10 $\sigma$) in step of 1 Jy~beam$^{-1}$. All the contours in the lower panel are started from \textbf{0.3 (3 $\sigma$),} 1 Jy~beam$^{-1}$ in step of 1 Jy~beam$^{-1}$. The dashed straight lines in the upper panel represent the contaminated molecular lines (SO$_{2}$ (13$_{3,11}$-13$_{2,12}$), OCS (22-21) \& HCN v2=1 (3-2) ). \textbf{These dashed straight lines are oblique due to velocity gradient.} The red and blue crosses in the lower panel mark the peak velocities of SO$_{2}$ (28$_{4,24}$-28$_{3,25}$) emission lines and HCO$^{+}$ (3-2) absorption lines at each position from Gaussian fits. }
\end{figure}

\begin{figure}
\centering
\includegraphics[angle=0,scale=0.5]{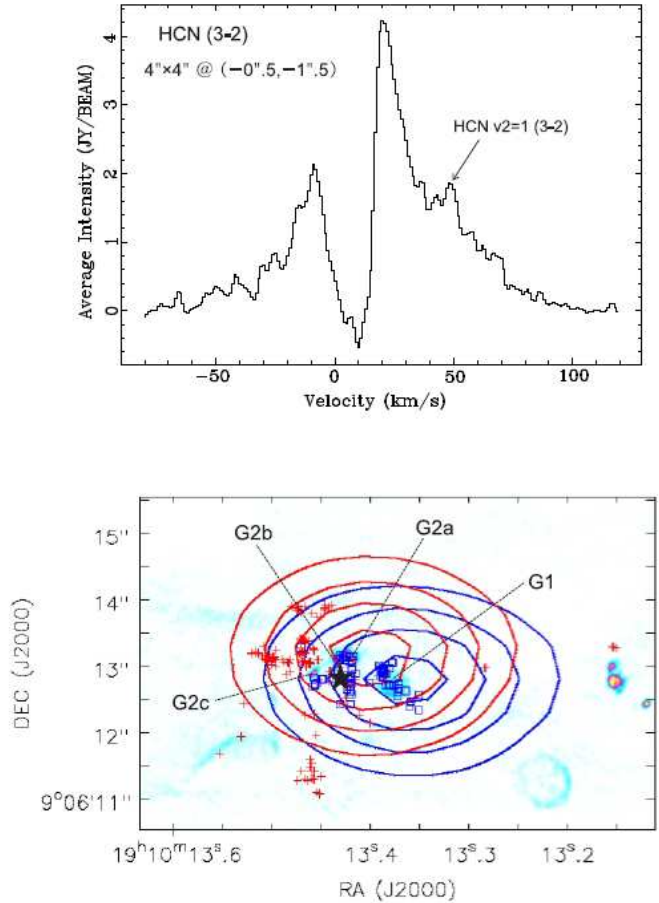}
\caption{Upper: Spectrum of HCN (3-2) line, which is averaged over 4$\arcsec\times4\arcsec$ area centered at position (-0$\arcsec.5$,-1$\arcsec.5$) offset from the phase reference center. Lower: The 7 mm continuum emission \citep{de00} is shown in color image. The integrated intensity maps of high velocity blueshifted and redshifted emission of HCN (3-2) are shown in blue and red contours. The red crosses and blue boxes mark the positions of water masers with positive and negative LSR velocities, respectively. All the contours are started from 20\% in step of 20\% of the peak values. The peak values for the blue and red contours are 30, and 49.4 Jy~beam$^{-1}$~km~s$^{-1}$, respectively. The mid-IR source ``G:IRS1" is marked with yellow filled star. }
\end{figure}

\begin{figure}
\centering
\includegraphics[angle=0,width=110mm]{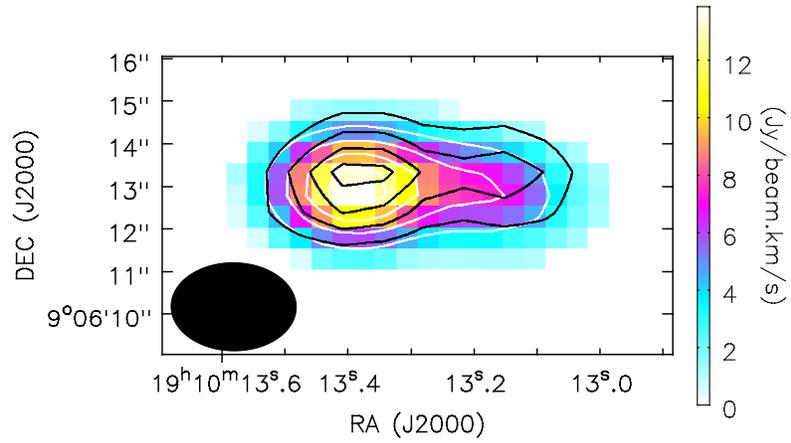}
\caption{The integrated intensity of HCN v2=1 (3-2) line is shown as white contours and color image. The 1.1 mm continuum emission is shown in black contours. All the contours are started from 20\% in step of 20\% of the peak values. The peak value of HCN v2=1 (3-2) is 13.9 Jy~beam$^{-1}$~km~s$^{-1}$. }
\end{figure}

\begin{figure}
\centering
\includegraphics[angle=0,width=130mm]{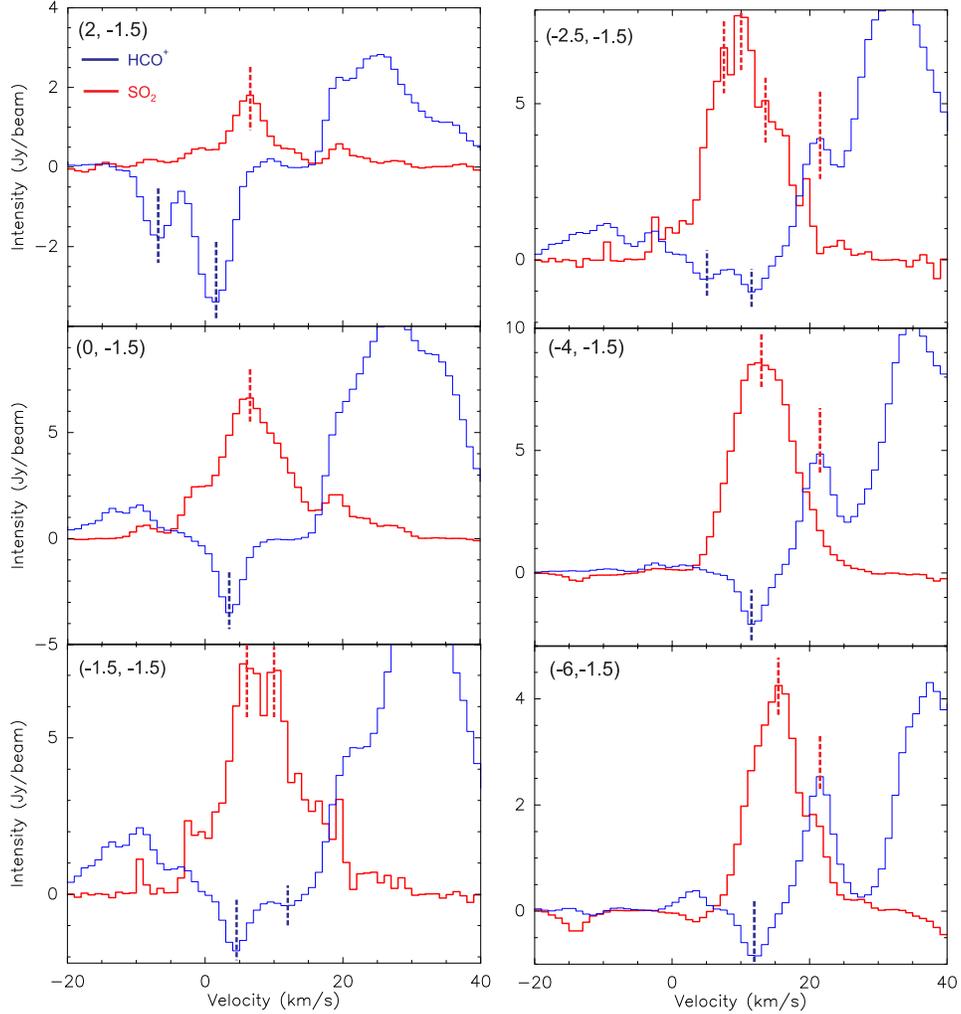}
\caption{Spectra of HCO$^{+}$ (3-2) and SO$_{2}$ (28$_{4,24}$-28$_{3,25}$) at six positions (labeled at the upper-right corner of each panel in unit of arcsec, \textbf{and take along the line shown in Figure2-bottom.} ) are shown in \textbf{red and blue}, respectively. The velocities of the absorption dips of HCO$^{+}$ (3-2) and emission peaks of HCO$^{+}$ (3-2) and SO$_{2}$ (28$_{4,24}$-28$_{3,25}$) are marked with \textbf{red and blue} dashed vertical lines, respectively.}
\end{figure}

\begin{figure}
\centering
\includegraphics[angle=0,width=80mm]{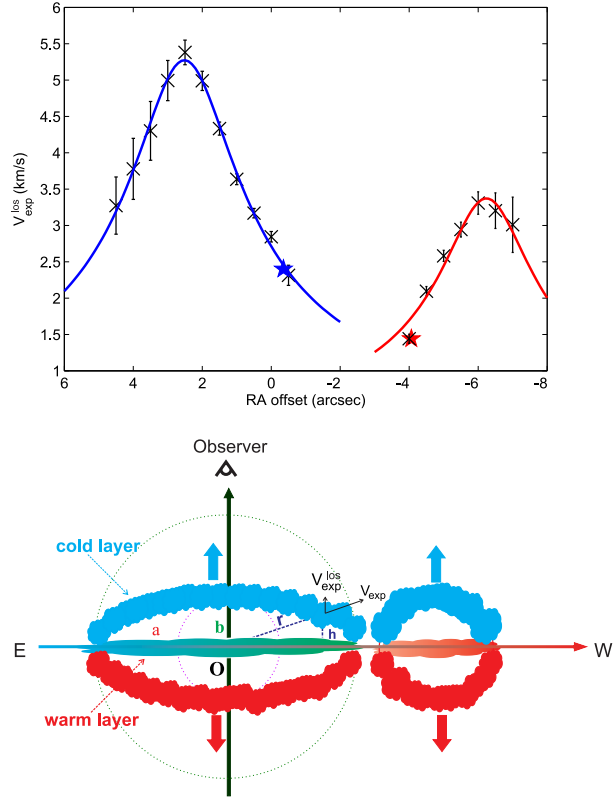}
\caption{Lower: A cartoon of one possible geometric model for causing the absorption in HCO$^{+}$ (3-2) lines in W49N. \textbf{There are two expanding systems. The cold gas traced by HCO$^{+}$ (3-2) absorption is distributed in two parallel expanding elliptical layers. The warm gas traced by SO$_{2}$ (28$_{4,24}$-28$_{3,25}$) is distributed along the middle plane. The color change of the horizon axis reflects the velocity gradient in SO$_{2}$ (28$_{4,24}$-28$_{3,25}$) emission. } The maximum height (``h") of the cold layer is ``b", and the maximum distance (``r") of the cold gas to the geometric center (``O") of the warm layer is ``a". The RA offset of the geometric center is ``X0". The maximum expanding velocity is ``V$_{max}$". Upper: The observed line-of-sight (los) expanding velocities at individual positions are plotted as crosses with errorbars. The line-of-sight (los) expansion velocity V$_{exp}^{los}$ is defined as V$_{exp}^{los}$=V$_{emi}$-V$_{abs}$, where V$_{emi}$ is the velocity of emission peak of SO$_{2}$ (28$_{4,24}$-28$_{3,25}$) and V$_{abs}$ is the velocity of absorption dip of HCO$^{+}$ (3-2). The two 1.1 mm continuum emission peaks are shown in stars. The best fits to the position-velocity curves are shown in red and blue solid lines, which correspond to [b, a, X0, V$_{max}$] of [1.18$\arcsec$, 2.84$\arcsec$, -6.24$\arcsec$, 3.37 km~s$^{-1}$] and [1.43$\arcsec$, 4.57$\arcsec$, 2.52$\arcsec$, 5.27 km~s$^{-1}$], respectively.  }
\end{figure}

\end{document}